\documentclass[a4paper,preprintnumbers,showpacs,twocolumn,superscriptaddress,nofootinbib,amsmath,amssymb]{revtex4-1}
\usepackage{color,hyperref}
\usepackage{times}
\usepackage{mathrsfs}
\hypersetup{colorlinks=true,linkcolor=blue,citecolor=magenta,filecolor=magenta,urlcolor=blue}

\usepackage{enumitem}
\usepackage{graphicx,subfigure}
\usepackage{dcolumn}
\usepackage{bm}
\usepackage{color}

\def\prd{Phys. Rev. D}

\def\apjl{Astrophysical Journal Letters}

\def\nat{Nature}

\providecommand{\dif}{\mathrm{d}} 

\newcommand{\beq}{\begin{equation}}
\newcommand{\eeq}{\end{equation}}
\newcommand{\bea}{\begin{eqnarray}}
\newcommand{\eea}{\end{eqnarray}}

\providecommand{\dif}{\mathrm{d}} \def\d{\dif}

\graphicspath{{pic/}}

\def\Opava{Research Centre for Theoretical Physics and Astrophysics, Institute of Physics, Silesian University in Opava, CZ-74601 Opava, Czech Republic}

\def\newuu{New Uzbekistan University, Movarounnahr str. 1, Tashkent 100000, Uzbekistan}
\def\astro{Ulugh Beg Astronomical Institute, Astronomicheskaya 33, Tashkent 100052, Uzbekistan}

\begin{document}

\title{Effect of nonlinear electrodynamics on polarization distribution around black hole}

\author{Abylaikhan Tlemissov}
\email{tlemissov-ozzy@mail.ru}
\affiliation{\Opava}
\author{Bobir Toshmatov}
\email{toshmatov@astrin.uz}
\affiliation{\newuu}
\affiliation{\astro}
\author{Ji\v{r}\'{i} Kov\'{a}\v{r}}
\email{jiri.kovar@physics.slu.cz}
\affiliation{\Opava}

\begin{abstract}
We have investigated the polarized images of synchrotron emission from magnetically charged, spherically symmetric regular Bronnikov black hole (in general relativity coupled to nonlinear electrodynamics) and the singular Reissner-Nordstr\"{o}m black hole (in general relativity coupled to Maxwell electrodynamics). By taking into account the fact that within the framework of nonlinear electrodynamics, the propagation of the light ray is governed not by the null geodesics of the original spacetime, but rather by the null geodesics associated with the effective metric, we have examined synchrotron radiation characteristics near these black holes at different inclination angles. Our calculations reveal that intensity differences between these two black holes are significant, especially at high values of magnetic charges and large tilt angles. Polarization quantities, such as the electric vector polarization angle ($\Delta$EVPA) and normalized intensity difference ($\Delta I/I$) show similar trends, driven by lensing, Doppler effects, and redshift, offering insights into black hole metrics.
\end{abstract}

\maketitle
\section{Introduction}

Over the last decades, there have been received numerous observational data that allow us to observe and investigate near-relativistic processes. Some of the most significant such data are the images of M87 \cite{EHT:2019dse} and Sgr A* \cite{EHT:2022wkp} and near-infrared (NIR) flares captured by GRAVITY \cite{GRAVITY:2020lpa}. These observations have opened many windows for testing the laws and alternative theories of gravity. One of the sources of such data is the Event Horizon Telescope (EHT) collaborations that have provided breakthrough images of the shadows of supermassive black holes, particularly Sgr A*, located at the center of our galaxy, and M87, in the nearby Virgo cluster. These observations have offered direct visual evidence of the strong-field regime of black holes. The shadow images confirm theoretical predictions regarding the photon orbit and the bending of light due to the immense gravitational field. Meanwhile, the GRAVITY collaboration has revealed NIR flares in the vicinity of Sgr A*, which are believed to originate from hotspots orbiting close to the black hole's event horizon. These observations provide insights into the accretion dynamics and the plasma physics in extreme conditions. By comparing these results with theoretical models, physicists can test general relativity with precision and explore the applicability of alternative theories of gravity. 

The plasma medium around a black hole can generate magnetic fields through the motion of charged particles in the accretion disk, creating electric currents that amplify the magnetic fields. These magnetic fields are essential for jet formation and are evidenced by polarized synchrotron radiation observations near black holes. From this perspective, in the paper \cite{EHT:2021btj,2021PhRvD.104d4060G}, a basic model for comprehending polarized black hole images is created. In this model, a ring of charged fluid circles a Schwarzschild black hole in the background magnetic field and equatorial fluid velocity, observer inclination, magnetic field profile, and emission radius can be arbitrarily chosen. The ring model provided a comprehensive representation of the overall morphology of the image obtained by the EHT observation \cite{2022EPJC...82..835Z}. Despite the ring model's ability to accurately present certain properties, it exhibits significant drawbacks, including depolarization problem. This discrepancy is due to the fact that when we obtain a polarized image by the EHT, we observe a fractional polarization with low resolution due to the internal Faraday effects. However, for the NIR flares \cite{2003Natur.425..934G}, the Faraday effect, absorption, and background radiation have negligible impact, allowing for a direct comparison between the observed polarization values and centroid motion with the simulated model for hot spots alone.

Although general relativity has demonstrated significant successes in the weak gravitational fields regime through solar system tests \cite {Will:2014kxa} and in the strong gravitational fields regime through the detection of gravitational waves by the LIGO and Virgo collaborations \cite{GW151226,GW170104,GW170608,GW170814,GW170817}, as well as the images of supermassive black holes captured by the EHT collaborations, there remains an opportunity for discourse regarding the issue of singularity at the center of the black hole within general relativity. It is a well-known fact that black holes in pure general relativity possess central curvature singularities that are unavoidable within the framework of the theory and cannot be adequately explained by general relativity alone. One of the approaches proposed to eliminate this problem is to couple general relativity to nonlinear electrodynamics \cite{Ayon-Beato:1998hmi,Bronnikov:2000yz}. So far, numerous regular spherically symmetric black hole solutions within this framework have been proposed and various phenomena associated with such black holes have been investigated -- see Refs. \cite{Dymnikova04,Novello01,Obukhov02,Lemos16,Breto16,Fan:PRD:2016,Toshmatov:2018ell,Nomura:2020tpc,Toshmatov:2021fgm,Moreno:2002gg} and references therein. Various studies have been conducted to explore the potential distinguishing characteristics between the regular black holes in general relativity coupled with nonlinear electrodynamics and the Reissner-Nordstr\"{o}m black holes in general relativity coupled with Maxwell (linear) electrodynamics \cite{2019ApJ87412S,Toshmatov:2019gxg,2019ApJ145S}. The polarization patterns of regular Bardeen, Hayward, and Schwarzschild black holes were analyzed in \cite{2022SCPMA..6520411L} using the same magnetic field profile. However, the authors did not reference an effective metric instead, the calculations were likely performed by using the standard metric, rather than employing an effective metric to describe the motion of light. In this paper, we shall examine the potential distinguishing properties by analyzing polarized images of a synchrotron-emitting ring surrounding Reissner-Nordstr\"{o}m and regular Bronnikov black holes. In our calculations, we assume that both black holes are magnetically charged, thus possessing their own magnetic fields.  

The paper is organized as follows. In section \ref{sec-background}, we briefly present the backgrounds of singular Reissner-Nordstr\"{o}m black holes in general relativity coupled to Maxwell electrodynamics and regular Bronnikov black holes in general relativity coupled to nonlinear electrodynamics. In section \ref{pol_vec_cal_section} we present a formalism for the polarization vector and gravitational lensing in the generic spherically symmetric black hole spacetime associated with the electrodynamics. In section \ref{EVPA_sec}, we elucidate the polarization angle of the electric vector for photons emitted in the vicinity of black holes. In the concluding sections \ref{sec-setup} and \ref{sec-conc}, we present the outcomes of our computational endeavors and provide an analysis of the main findings elucidated in this manuscript. Throughout the paper, we adopt geometric units where $G=1=c$ together with $1/(4\pi \epsilon_{0})=1$.

\section{Background and magnetic field profile in P, F-frame}\label{sec-background}

The action of general relativity coupled to electrodynamics is given by
\begin{equation}\label{action}
    S=\frac{1}{16\pi}\int d^4x \sqrt{-g}\left(R-L(F)\right),
\end{equation}
where $g$ denotes the determinant of the metric tensor, $R$ represents the Ricci scalar of the spacetime, and $L$ denotes the Lagrangian density of the electrodynamics, which is expressed as a function of the electromagnetic field invariant $F=F_{\mu\nu}F^{\mu\nu}$, with $F_{\mu\nu}=\partial_\mu A_\nu-\partial_\nu A_\mu$ representing the electromagnetic field tensor. By applying the least action principle for the action (\ref{action}), one obtains the Einstein and Maxwell equations which are given by
\begin{eqnarray}\label{field-eqs}
    &&G_{\mu\nu}=T_{\mu\nu}, \qquad T_{\mu\nu}=2\left(L_F F_{\mu\nu}^2-\frac{1}{4}g_{\mu\nu}L\right)\ ,\\
    &&\nabla_\mu\left(L_F F^{\mu\nu}\right)=0\ .
\end{eqnarray}
where $L_F=\partial_F L$. To fully establish the background, it is necessary to select a spacetime metric, for which we employ the spherically symmetric static line element as
\begin{equation}\label{line-element}
    ds^2=-f(r)dt^2+\frac{dr^2}{f(r)}+r^2\left(d\theta^2+\sin^2{\theta}d\phi^2\right)\ ,
\end{equation}
and the ansatz for the magnetically charged (with charge $q$) electromagnetic field as 
\begin{equation}
    A_{\mu}=-q\cos{\theta}\delta_\mu^\phi, \quad F_{23}=q\sin{\theta}.
\end{equation}
If the Lagrangian density of the electrodynamics is a linear function of $F$ ($L=F$), specifically in the context of Maxwell electrodynamics, then the resulting field equations (\ref{field-eqs}) yield the well-known Reissner-Nordstr\"{o}m black hole solution, for which the metric function is expressed as
\begin{equation}
    f_{\rm{RN}}(r)=1-\frac{2M}{r}+\frac{q^2}{r^2}
\end{equation}
Together with the Reissner-Nordstr\"{o}m black hole solution, in the current paper we examine the regular Bronnikov black hole solution which is obtained from the following model of the Lagrangian density of the nonlinear electrodynamics \cite{Bronnikov01}:
\begin{equation}
    L(F)=F \cosh^{-2}{\left(a\left|\frac{F}{2}\right|^{1/4}\right)}.
\end{equation}
The above model gives the metric function in the following form: 
\begin{equation}
    f(r)=1-\frac{|q|^{3/2}}{a r}\left(1-\tanh{\frac{a\sqrt{|q|}}{r}}\right).
    \label{laps_1}
\end{equation}
Where $q$ is a magnetic charge and $a$ is an integration constant. In order to simplify the metric function (\ref{laps_1}), we introduce the total mass of the spacetime in terms of the ADM (Arnowitt–Deser–Misner) mass which is determined from the asymptotic behavior of the metric function (\ref{laps_1}) as
\begin{eqnarray}\label{laps-asymp}
    f(r)=1-\frac{|q|^{3/2}}{a r}+O\left(\frac{1}{r^2}\right)\ .
\end{eqnarray}
From (\ref{laps-asymp}) one can easily notice that the total mass equals $M=|q|^{3/2}/2a$. Consequently, the metric function \eqref{laps_1} can be reformulated into that of the Schwarzschild black hole as
\begin{eqnarray}
    f(r)&=&1-\frac{2m(r)}{r},
    \label{lapse_function}
\end{eqnarray}
with mass function
\begin{eqnarray}
    m(r)&=&M\left(1-\tanh{\left(\frac{q^2}{2 M r}\right)}\right).
    \label{mass_function}
\end{eqnarray}
The solution presented herein exhibits asymptotic flatness at spatial infinity. Furthermore, at large distances, it emulates the characteristics of Reissner-Nordstr\"{o}m spacetime as
\begin{equation}    
f(r)=f_{\rm{RN}}(r)+O\left(\frac{1}{r^4}\right).
\end{equation}
It is noteworthy that the Reissner-Nordstr\"{o}m spacetime, a solution of the Einstein-Maxwell equations, is identical for both electrically and magnetically charged cases. As previously mentioned, the Bronnikov black hole solution tends to the Reissner-Nordstr\"{o}m solution in the weak gravitational field regime, which occurs when the black hole is either weakly charged or observed at large distances. However, in the strong field regime, these two solutions differ significantly. The most notable distinction between them lies in their event horizon radii. It is well-established that the presence of electric or magnetic charge in a black hole reduces its gravitational pull, leading to a decrease in the event horizon radius. The Reissner-Nordstr\"{o}m black hole can have a maximum charge $q_{\rm max}=M$ where its event horizon radius becomes $r_{\rm H, min}=M$. However, in the Bronnikov black hole $q_{\rm max}\approx 1.055M$, $r_{\rm H, min}\approx 0.87M$. Below in Fig. \ref{fig-horizon}, we present the radii of the event horizon of these black holes as a function of their charges.
\begin{figure*}[th]
\centering
\includegraphics[width=0.45\textwidth]{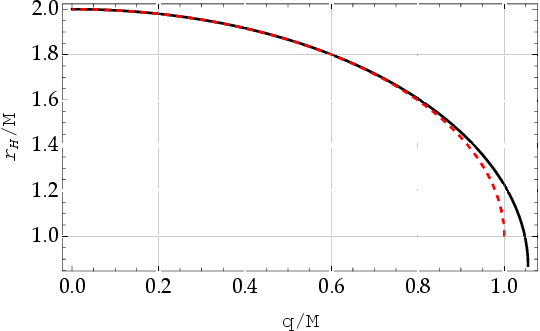}
\includegraphics[width=0.45\textwidth]{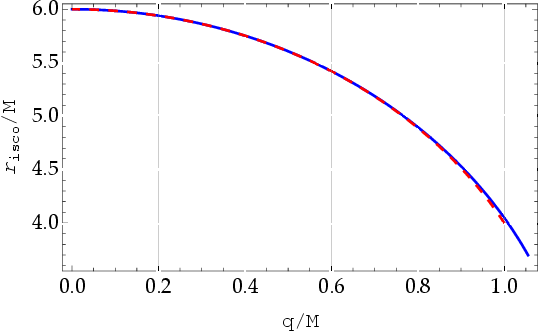}
\caption{Radii of event horizon (left panel) and ISCO (right panel) of the Bronnikov (blue curve) and Reissner-Nordstr\"{o}m (red dashed curve) black holes as a function of their charges.}
\label{fig-horizon}
\end{figure*}
Moreover, we aim to highlight another crucial radius known as the innermost stable circular orbit (ISCO), which marks the inner boundary of the region where stable circular orbits can exist. Given that our study assumes the light-emitting ring is stable, understanding the location of the stable circular orbits is of significant importance. On the right panel of Fig. \ref{fig-horizon}, we present the ISCO radii for both black holes examined in this paper. From this analysis, it is evident that for small to intermediate values of the black hole charge, the Bronnikov and Reissner-Nordstr\"{o}m black holes exhibit very similar event horizon and ISCO radii. However, for larger charge values, the Bronnikov black hole demonstrates stronger gravitational attraction. Thus, hereafter in our calculations we select radius of the light-emitting ring, $r_s$, is bigger than the ISCO radius.

Now let us construct an orthonormal frame of reference at the emitted point-P (P-frame) with the given local coordinates $x^{(a)}=\{\hat{t},\hat{x},\hat{y},\hat{z}\}$ located at $r=r_{s}$ in the equatorial plane, $\theta_{s}=\pi/2$. As soon as $\{t, r, \theta, \phi\}$ in global coordinates are parallel with $\{\hat{t}, \hat{x}, \hat{z}, \hat{y}\}$ axis in local coordinates, we use corresponding basis vectors $e_{(a)}=e^{\mu}_{(a)}\partial_{\mu}$.
\begin{eqnarray}\label{local basis_1}
    e_{(\hat{t})}&=&\frac{1}{\sqrt{f(r_{s})}}\partial_{t},\quad e_{(\hat{x})}=\sqrt{f(r_{s})}\partial_{r},\nonumber\\
    \label{local basis_2}
    e_{(\hat{y})}&=&\frac{1}{r_{s}}\partial_{\phi},\quad 
    e_{(\hat{z})}=\frac{1}{r_{s}}\partial_{\theta}.
\end{eqnarray}
The tetrads satisfy the condition $g_{\mu\nu}e^{\mu}_{(a)}e^{\nu}_{(b)}=\eta_{ab}$ and are constructed so that, in the local reference of frame, the spacetime is described by the Minkowski metric. One can transform 4-vectors and tensors from global coordinates to the local one through
\begin{eqnarray}
    V^{(a)}&=&\eta^{(a)(b)}e^{\mu}_{(b)}V_{\mu},\quad T^{(a)(b)}=\eta^{(a)(c)}\eta^{(b)(d)}e^{\mu}_{(c)}e^{\nu}_{(d)}T_{\mu\nu}\nonumber\\
    V_{(a)}&=&e^{\mu}_{(a)}V_{\mu},\quad T_{(a)(b)}=e^{\mu}_{(a)}e^{\nu}_{(b)}T_{\mu\nu}.
\end{eqnarray}
Therefore, the components of the electromagnetic field tensor in the local coordinates become
\begin{equation}    
    F_{\hat{2}\hat{3}}= e^{2}_{(\hat{2})}e^{3}_{(\hat{3})}F_{23}= \frac{q}{r_{s}^2},
\end{equation}
which means that the magnetic field is oriented along the $\hat{x}$ direction and explicitly is given by
\begin{equation}
    \vec{B}_{(P)}=\frac{q}{r_{s}^2}\hat{x}=B_{\hat{x}(P)}\hat{x}.
\end{equation}
Now suppose charged fluid rotates around black hole with nonzero velocity on $\hat{x}\hat{y}$ plane
\begin{equation}   
    \vec{\beta}= \beta\left(\cos{\chi}\hat{x}+\sin{\chi}\hat{y}\right).
\end{equation}
The sign of $\chi$ corresponds to clockwise $\chi>0$ or anti-clockwise $\chi<0$ rotation. Then, in order to calculate field components in the fluid frame (F-frame), we use the Lorentz transformations which lead \cite{EHT:2021btj}
\begin{eqnarray}  
    B_{\hat{x}(F)}&=&\left(\cos^2{\chi}+ \gamma\sin^2{\chi}\right)B_{\hat{x}(P)}\\
    B_{\hat{y} (F)}&=&-(\gamma-1)\cos{\chi}\sin{\chi}B_{\hat{x}(P)},
\end{eqnarray}
where the Lorentz factor $\gamma=\frac{1}{\sqrt{1-\beta^2}}$ is defined in terms of the velocity $\beta$ of a charged fluid. 

\section{polarization vector and propagation of light}\label{pol_vec_cal_section}

In this section, we consider the polarization vector, which is defined in $F-$ frame as a vector, perpendicular to the magnetic field $\vec{f}\perp \vec{B}$ and momentum $\vec{f}\perp \vec{k}$ direction. From the above, it follows that:
\begin{equation}  
    \vec{f}_{(F)}=\frac{\left(\vec{k}_{F}\times \vec{B}_{F}\right)}{|\vec{k}_{F}|}, \quad f_{t(F)}=0.
\end{equation}
By the definition, polarization vector is normalized as $f^{\mu}f_{\mu}=\sin^2{\zeta}|\vec{B}_{(F)}|$ where $\zeta$ is an angle between magnetic field and momentum, which can be calculated as follows:
\begin{equation}
    \sin{\zeta}= \frac{\big|\left(\vec{k}_{F}\times \vec{B}_{F}\right)\big|}{|\vec{k}_{F}||\vec{B}_{F}|}.
\end{equation}

\begin{figure}[th]
\centering
\includegraphics[width=0.45\textwidth]{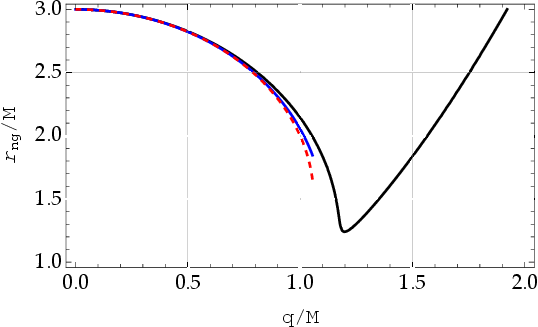}
\caption{The radii of the photonsphere of the Reissner-Nordstr\"{o}m (red dashed curve) and the Bronnikov (black curve) black holes as well as the circular null geodesics of the Bronnikov (blue curve) spacetime as a function of charge of the black hole.}
\label{fig-ps}
\end{figure}

In the P-frame, the polarization vector must be calculated via $f^{(a)}_{(P)}=\Lambda^{\hspace{3mm}(a)}_{(b)}f_{(F)}^{(b)}$ inverse Lorentz boost, which yields
\begin{eqnarray}
    f_{(P)}^{\hat{t}}&=&\gamma f^{\hat{t}}_{(F)}+\gamma \beta \cos{\chi}f^{\hat{x}}_{(F)}+\gamma \beta \sin{\chi}f^{\hat{y}}_{(F)}\\
    f_{(P)}^{\hat{x}}&=&\gamma\beta \cos{\chi} f^{\hat{t}}_{(F)}+(1+\left(\gamma-1\right)\cos^2{\chi})f_{(F)}^{\hat{x}}\\
    & &+(\gamma-1)\cos{\chi}\sin{\chi}f_{(F)}^{\hat{y}}\nonumber\\
    f_{(P)}^{\hat{y}}&=&\gamma\beta \sin{\chi} f^{\hat{t}}_{(F)}+(\gamma-1)\cos{\chi}\sin{\chi}f_{(F)}^{\hat{x}}\\
    & &+(1+\left(\gamma-1\right)\sin^2{\chi})f_{(F)}^{\hat{y}}\nonumber\\
    f_{(P)}^{\hat{z}}&=&f_{(F)}^{\hat{z}}
\end{eqnarray}
It is crucial to note that, within the framework of nonlinear electrodynamics, the propagation of the light ray is governed not by the null geodesics of the original spacetime, but rather by the null geodesics associated with the effective metric \cite{Novello:1999pg,Novello:2000km} and it appears just as a convenient computational element for the light ray. It is important to note that the motion of a massive particle is governed by the original physical line element (\ref{line-element}). Now let's consider the propagation of light from the point of view of a photon moving in an effective metric
\begin{equation}\label{eff-line-element}
    ds_{\rm{eff}}^2=-\frac{f(r)}{L_{\rm{F}}}dt^2+ \frac{1}{L_{\rm{F}}f(r)}dr^2+\frac{r^2}{\Phi}\left(d\theta^2+\sin^2{\theta}d\phi^2\right)\ ,
\end{equation}
where $\Phi=L_{\rm{F}}+2FL_{\rm{FF}}$ with $L_{\rm{F}}$, $L_{\rm{FF}}$ being the first ($L_F=\partial_F L$) and second ($L_{FF}=\partial_{F}^2 L$) derivatives of the Lagrangian density with respect to $F$, respectively. Although it is more convenient to express $L_{\rm{F}}$, $L_{\rm{FF}}$ and $\Phi$ in terms of the mass function \eqref{mass_function} by using $F=2q^2/r^4$ which gives
\begin{eqnarray}  
    L_{\rm{F}}(r)&=&\frac{r^2\left(2m'-rm''\right)}{2 q^2}\ ,\\
    L_{\rm{FF}}(r)&=&\frac{r^6\left(-4m'+rm''+r^2m'''\right)}{16 q^4}\ ,\\
    \Phi(r)&=&\frac{r^3\left(rm'''-m''\right)}{4 q^2}\ .
\end{eqnarray}
As illustrated by the event horizon and ISCO radii in Fig. \ref{fig-horizon}, the Bronnikov and Reissner-Nordstr\"{o}m black holes exhibit very similar values for these radii. Another fundamental radius of great relevance to our current calculations is the photon sphere radius. As previously discussed in this section, in spacetimes unrelated to nonlinear electrodynamics, light rays follow the null geodesics of the spacetime, and the photon sphere corresponds to the circular null geodesics of the original spacetime (\ref{line-element}). However, for the Bronnikov black hole, the photon sphere is determined by the circular null geodesics of the effective spacetime metric (\ref{eff-line-element}). In Fig. \ref{fig-ps}, we present the photon sphere radii for both the Reissner-Nordstr\"{o}m and Bronnikov black holes. Additionally, we include the radii of the circular null geodesics for the Bronnikov black hole to highlight the influence of the effective metric. 
One can see from this figure that the Bronnikov black hole (or no-horizon spacetime when $q>1.055M$) always has a nonvanishing photon sphere, as it is a unique property of the black holes within general relativity coupled to the nonlinear electrodynamics \cite{Toshmatov:2019gxg}.

\begin{figure}[th]
\centering
\includegraphics[width=0.45\textwidth]{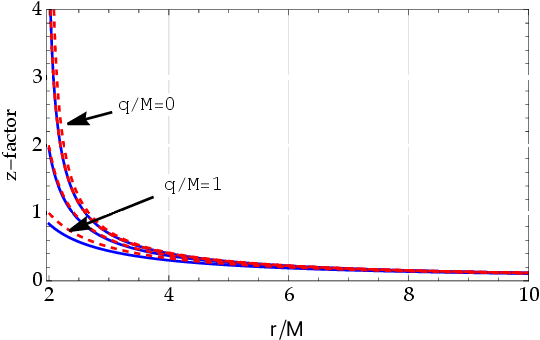}
\caption{The $z$-factor in the Bronnikov (blue curve) and Reissner-Nordstr\"{o}m (red dashed curve) black holes as a function of radial coordinate. Where charge ranges in $q/M\in[0,1]$ with increments of 0.33.}
\label{fig-z-factor}
\end{figure}

According to \cite{Bronnikov01,2000CQGra..17.3821N} the redshift factor from emitter to observer in the case of a magnetically charged black hole becomes
\begin{equation}
    1+z=\frac{L_{F}}{\sqrt{f(r)}}\ .
\end{equation}
In Fig. \ref{fig-z-factor} we present the $z$-factor as a function of radial coordinate in the field of the Bronnikov and Reissner-Nordstr\"{o}m black holes and it shows that the $z$-factor for both spacetimes is almost identical and it dissipates very rapidly with increasing the radial coordinate.

Now we consider the photon with normalized energy $k_{t}=-1$ moving from emitter to observer. It is evident that, in the context of spherical symmetric spacetime, the emitted photon is associated with the $\hat{x}$$\hat{z}$ plane in the geodesic frame \cite{EHT:2021btj,2022EPJC...82..784Q}. This observation leads to the following relations for photon momentum in (G)-frame.
\begin{eqnarray}
    k^{\hat{t}}_{(G)}&=&\frac{L_{\rm{F}}(r_{s})}{\sqrt{f(r_{s})}}, \quad k^{\hat{y}}_{(G)}=0, \nonumber\\
    k_{(G)}^{\hat{x}}&=&k^{t}_{(G)}\cos{\alpha},\quad k_{(G)}^{\hat{z}}=k^{t}_{(G)}\sin{\alpha}
\end{eqnarray}
where we applied the redshift factor calculated by using the effective metric. Therefore, corresponding orthonormal components in (P)-frame take the form
\begin{eqnarray}
    k^{\hat{t}}_{(P)}&=&\frac{L_{\rm{F}}(r_{s})}{\sqrt{f(r_{s})}}, \quad k^{\hat{y}}_{(P)}=-\frac{L_{\rm{F}}(r_{s})\sin{\xi}\sin{\alpha}}{\sqrt{f(r_{s})}},\nonumber\\
    k_{(P)}^{\hat{x}}&=&\frac{L_{\rm{F}}(r_{s})\cos{\alpha}}{\sqrt{f(r_{s})}}, \quad k_{(P)}^{\hat{z}}=\frac{L_{\rm{F}}(r_{s})\cos{\xi}\sin{\alpha}}{\sqrt{f(r_{s})}}.
\end{eqnarray} 
Here $\alpha-$ emission angle, $\xi-$ satisfies relations
\begin{equation}
    \cos{\xi}=\frac{\cos{\theta_{0}}}{\sin{\psi}},\qquad \sin{\xi}=\frac{\sin{\theta_{0}}\cos{\phi}}{\sin{\psi}},
\end{equation}
where $\psi-$ angle between $\hat{x}$ and unit vector $\hat{n}$ towards the observer. At the next step, one must rewrite the emission angle $\alpha$ in terms of $\psi$. To do so, we calculate the angle $\psi$ by assuming the observer is located at infinity as
\begin{equation}
    \psi=\int_{r_{s}}^{\infty}\frac{\Phi \d r}{r^2\sqrt{\frac{L_{F}^2}{b^2}-\frac{f\Phi}{L_{F}r^2}}}\ ,
\end{equation}
together with
\begin{equation}
    \sin{\alpha}=\frac{b}{r_{s}L_{F}(r_{s})}\sqrt{\frac{f(r_{s})\Phi(r_{s})}{L_{F}(r_{s})}}
    \label{sin_rel}
\end{equation}
which helps us to find the relationship between $\psi$ and $\alpha$ emission angle. The relation \eqref{sin_rel} is found by using the fact $\tan{\alpha}=\sqrt{u^{\psi}u_{\psi}}/\sqrt{u^{r}u_{r}}$ (for more detail see \cite{2002ApJ...566L..85B}). Finally, we obtain the 4-momentum of the photon in the fluid frame by using the Lorentz transformations as
\begin{eqnarray}
    k_{(F)}^{\hat{t}}&=&\gamma k_{P}^{\hat{t}}-\gamma \beta \cos{\chi}k^{\hat{x}}_{P}-\gamma \beta \sin{\chi}k^{\hat{y}}_{(P)}\\
     k_{(F)}^{\hat{x}}&=&-\gamma \beta \cos{\chi}k^{\hat{t}}_{P}+\left(1+(\gamma-1)\cos^2{\chi}\right)k^{\hat{x}}_{(P)}\\
     & &+(\gamma-1)\cos{\chi}\sin{\chi}k_{(P)}^{\hat{y}}\nonumber\\
     k_{(F)}^{\hat{y}}&=&-\gamma \beta \sin{\chi}k^{\hat{t}}_{P}+(\gamma-1)\cos{\chi}\sin{\chi}k_{(P)}^{\hat{x}}\\
     & &+\left(1+(\gamma-1)\cos^2{\chi}\right)k^{\hat{y}}_{(P)}\nonumber\\
     k_{(F)}^{\hat{z}}&=&k_{(P)}^{\hat{z}}.
\end{eqnarray}
We must acknowledge that aforementioned formalism is applicable to any spherically symmetric black hole within the framework of general relativity associated with electrodynamics. Particularly, for the Reissner-Nordstr\"{o}m black hole $L_F=1$, $L_{FF}=0$, and $\Phi=1$.

\begin{figure*}[t!]
\centering
\includegraphics[width=.24\linewidth]{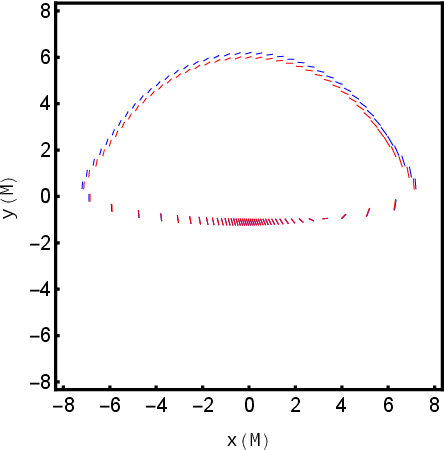}
\includegraphics[width=.24\linewidth]{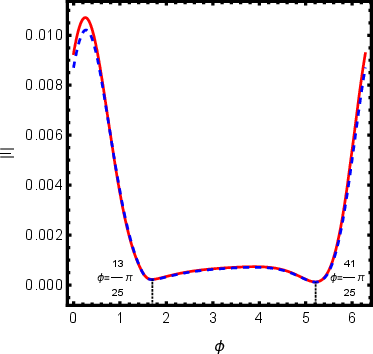} 
\includegraphics[width=.24\linewidth]{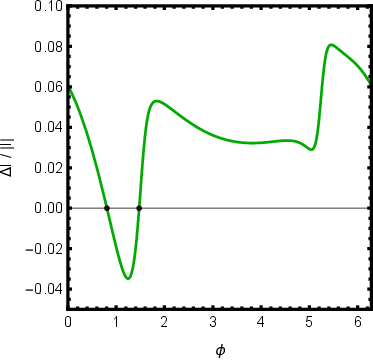} 
\includegraphics[width=.24\linewidth]{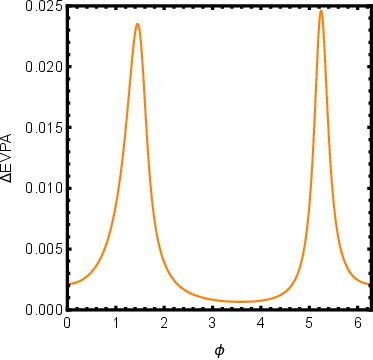}
\includegraphics[width=.24\linewidth]{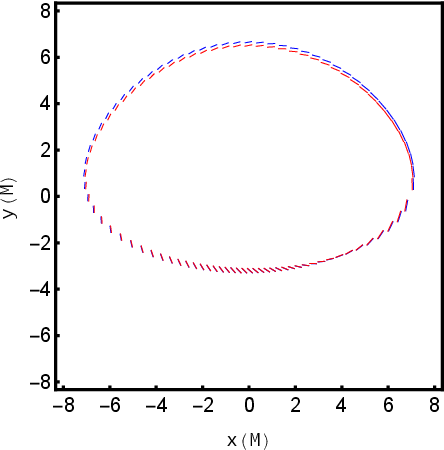}
\includegraphics[width=.24\linewidth]{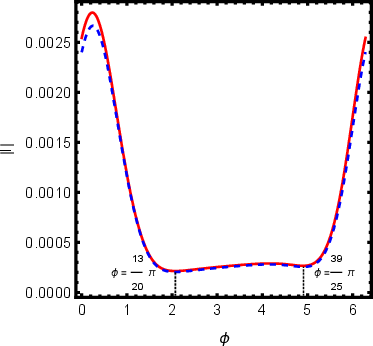} 
\includegraphics[width=.24\linewidth]{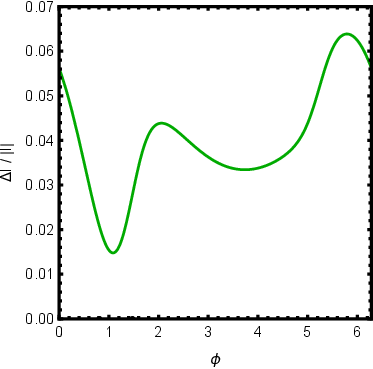} 
\includegraphics[width=.24\linewidth]{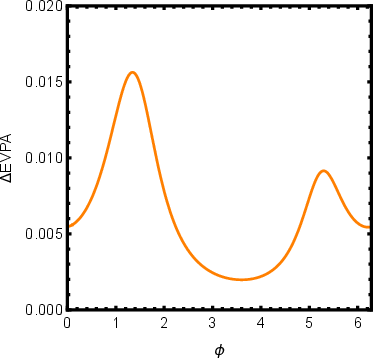}
\includegraphics[width=.24\linewidth]{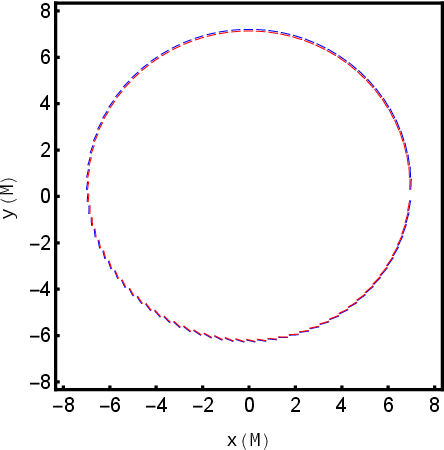}
\includegraphics[width=.24\linewidth]{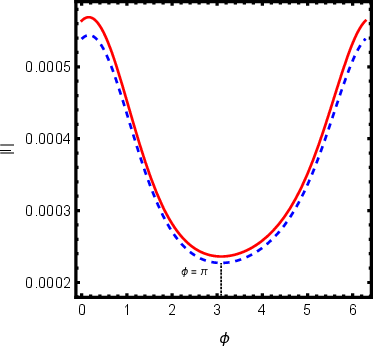} 
\includegraphics[width=.24\linewidth]{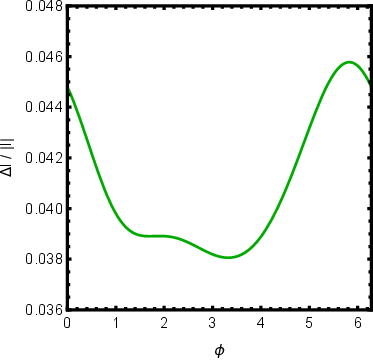} 
\includegraphics[width=.24\linewidth]{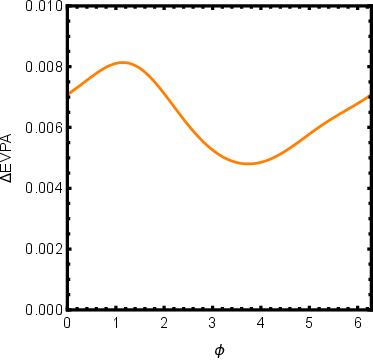}
\caption{Polarization image around regular black hole on observers screen, intensity, normalized intensity difference $\Delta I=(I-I_{\rm{RN}})/I$ and $\Delta EVPA=EVPA-EVPA_{\rm{RN}}$ in case of $q=0.99M$, $r_{s}=6M$ with different tilted axes $\theta_{0}=80^{\circ}$ (top panels), $\theta_{0}=60^\circ$ (middle panels) and $\theta_{0}=20^{\circ}$ (lower panels). The blue and red curves in polarimetric image and intensity profile correspond to the direction of polarization in the case of the Reissner-Nordstr\"{o}m and Bronnikov black holes, respectively.}
\label{equator_pic}
\end{figure*}

\begin{figure*}[t!]
\centering
\includegraphics[width=.24\linewidth]{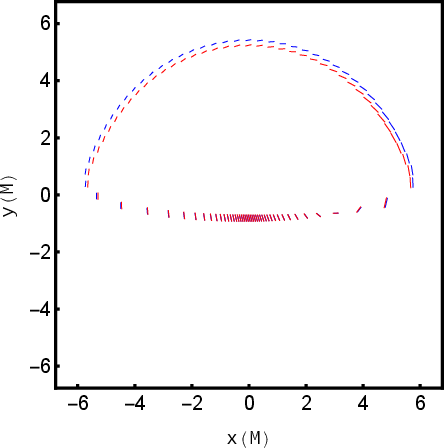}
\includegraphics[width=.24\linewidth]{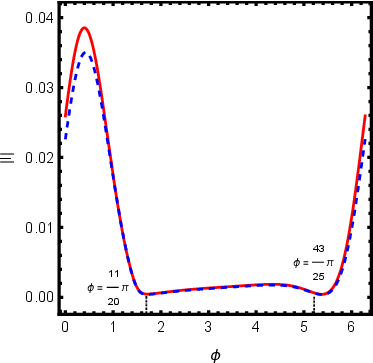} 
\includegraphics[width=.24\linewidth]{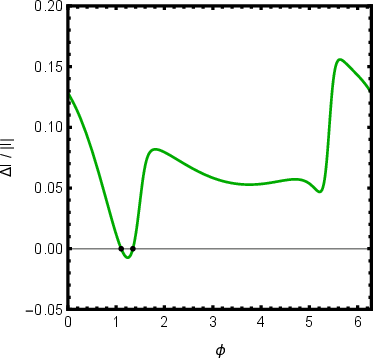} 
\includegraphics[width=.24\linewidth]{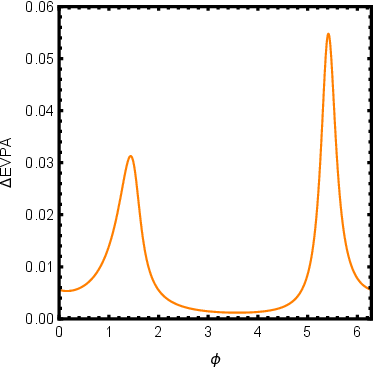}
\includegraphics[width=.24\linewidth]{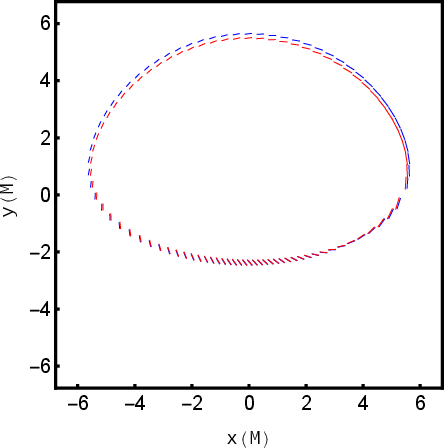}
\includegraphics[width=.24\linewidth]{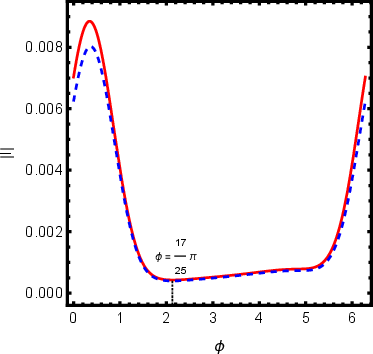} 
\includegraphics[width=.24\linewidth]{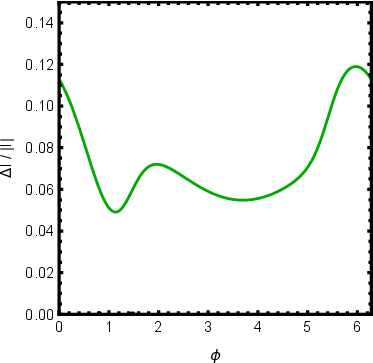} 
\includegraphics[width=.24\linewidth]{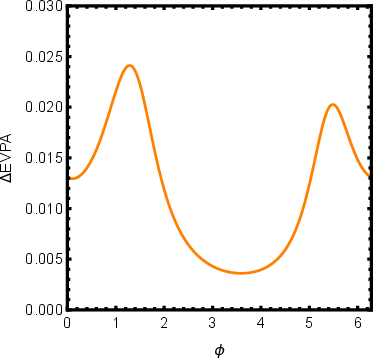}
\includegraphics[width=.24\linewidth]{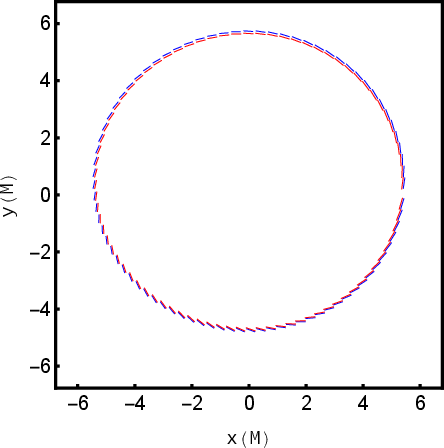}
\includegraphics[width=.24\linewidth]{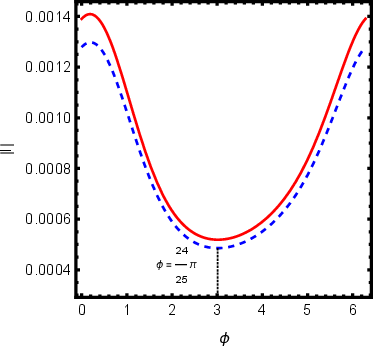}  
\includegraphics[width=.24\linewidth]{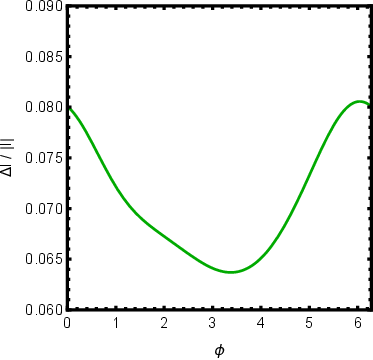} 
\includegraphics[width=.24\linewidth]{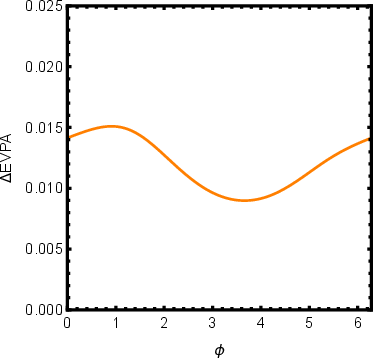}
\caption{ The same as Fig. \ref{equator_pic}, but for the values $q=0.99M$, $r_{s}=4.5M$.}
\label{equator_pic_q_099_rs_4_5}
\end{figure*}

\section{screen coordinates and EVPA}\label{EVPA_sec}

Now we are ready to introduce complex Penrose-Walker constant \cite{1970CMaPh..18..265W}
\begin{eqnarray}  
    \kappa&=&\kappa_{1}+i\kappa_{2},\\
    \kappa_{1}&=&\psi_{2}^{-1/3}\left(k^{t}f^{r}-k^{r}f^{t}\right),\\ \kappa_{2}&=&\psi_{2}^{-1/3}r_{s}^2\left(k^{\phi}f^{\theta}-k^{\theta}f^{\phi}\right),\\\nonumber
\end{eqnarray}
where we have momentum and polarization written in the Boyer-Lindquist coordinates, which is an inverse transformation from the local frame
\begin{eqnarray}
    k^{t}&=&\frac{k^{\hat{x}}_{(P)}}{\sqrt{f(r_{s})}},\quad k^{r}=\sqrt{f(r_{s})}k^{\hat{x}}_{(P)},\nonumber\\
    k^{\phi}&=&\frac{k^{\hat{y}}_{(P)}}{r_{s}},\quad k^{\theta}=\frac{k^{\hat{z}}_{(P)}}{r_{s}}
\end{eqnarray}
as well as for the polarization vector
\begin{eqnarray}
    f^{t}&=&\frac{f^{\hat{x}}_{(P)}}{\sqrt{f(r_{s})}},\quad f^{r}=\sqrt{f(r_{s})}f^{\hat{x}}_{(P)},\nonumber\\
    f^{\phi}&=&\frac{f^{\hat{y}}_{(P)}}{r_{s}},\quad f^{\theta}=\frac{f^{\hat{z}}_{(P)}}{r_{s}}.
\end{eqnarray}
together with Weyl scalar
\begin{equation}
    \psi_{2}=\frac{M}{r_{s}^{3}}\left(1-\tanh{\frac{q^2}{2 M r_{s}}}-\frac{q^2}{2M r_{s}}\times\frac{1}{\cosh{\frac{q^2}{2 M r_{s}}}}\right).
\end{equation}
The celestial coordinates $(x,y)$ for a photon moving towards the observer at infinity are described as follows:
\begin{eqnarray}
    x&=&-\frac{r_{s}k^{\hat{y}}_{(P)}}{\sin{\theta}},\\
    y&=&r_{s}\sqrt{\left(k^{\hat{z}}_{(P)}\right)^2-\left(k^{\hat{y}}_{(P)}\right)^2\cot^2\theta}sgn\left(\sin{\phi}\right).
\end{eqnarray}
The expression for the normalized electric field $\vec{E}$ associated with synchrotron radiation, as it propagates along a geodesic in the observer's sky coordinates, is given as follows:
\begin{eqnarray}
    E_{x,norm}&=&\frac{y\kappa_{2}+x\kappa_{1}}{\sqrt{\left(\kappa_{1}+\kappa_{2}\right)\left(x^2+y^2\right)}}\\
    E_{y,norm}&=&\frac{y\kappa_{1}-x\kappa_{2}}{\sqrt{\left(\kappa_{1}+\kappa_{2}\right)\left(x^2+y^2\right)}}\\
    E_{x,norm}^2&+&E_{y,norm}^2=1.
\end{eqnarray}
The components of the observed electric field and the intensity of synchrotron radiation, as received by an observer at infinity, are dependent on multiple factors and are expressed as follows:
\begin{eqnarray}    
    E_{x,obs}&=&\delta^{\frac{3+\alpha_{\nu}}{2}}l_{p}^{1/2}|\vec{B}|^{\frac{1+\alpha_{\nu}}{2}}\sin^{\frac{1+\alpha_{\nu}}{2}}{\zeta}E_{x,norm},\\
    E_{y,obs}&=&\delta^{\frac{3+\alpha_{\nu}}{2}}l_{p}^{1/2}|\vec{B}|^{\frac{1+\alpha_{\nu}}{2}}\sin^{\frac{1+\alpha_{\nu}}{2}}{\zeta}E_{y,norm},\\
    E_{x,obs}^2&+&E_{y,obs}^2= \delta^{3+\alpha_{\nu}}l_{p}|\vec{B}|^{1+\alpha_{\nu}}\sin^{1+\alpha_{\nu}}{\zeta},
\end{eqnarray}
where $\delta$ is the Doppler shift factor and $\alpha_{\nu}$ relates photon energy to disk temperature. The other quantity $l_{p}$ describes photon path length which is usually written in the form
\begin{equation}   
    l_{p}=\frac{k^{\hat{t}}_{(F)}}{k^{\hat{z}}_{(F)}}H,
\end{equation}
where $H$ is the height of the disk which can be chosen as a constant for simplicity. In the following, we set $\alpha_{\nu}=1$ for which
\begin{equation}                
    |I|=\delta^{4}l_{p}|\vec{B}|^2\sin^2{\zeta}\ , \quad EVPA=\frac{1}{2}\arctan{\frac{U}{Q}}\ ,
\end{equation}
where Stokes parameters $Q$ and $U$ are defined as
\begin{equation}
    Q=E_{y,obs}^2-E_{x,obs}^2\ , \quad U=-2E_{x,obs}E_{y,obs}.
\end{equation}
\begin{figure*}[t!]
\centering
\includegraphics[width=.24\linewidth]{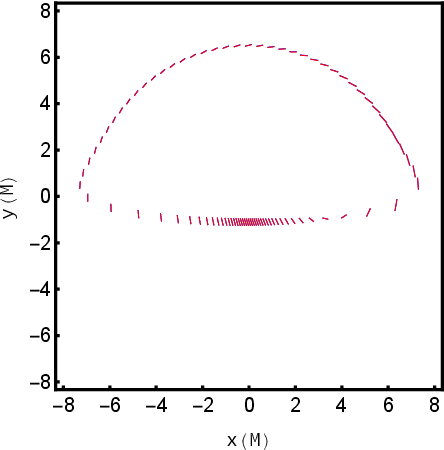}
\includegraphics[width=.24\linewidth]{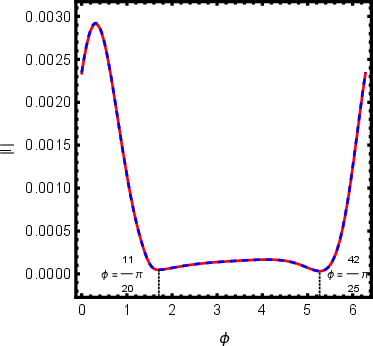} 
\includegraphics[width=.24\linewidth]{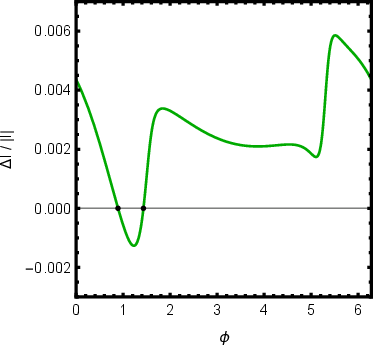} 
\includegraphics[width=.24\linewidth]{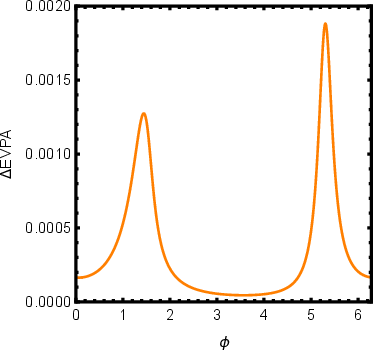}
\includegraphics[width=.24\linewidth]{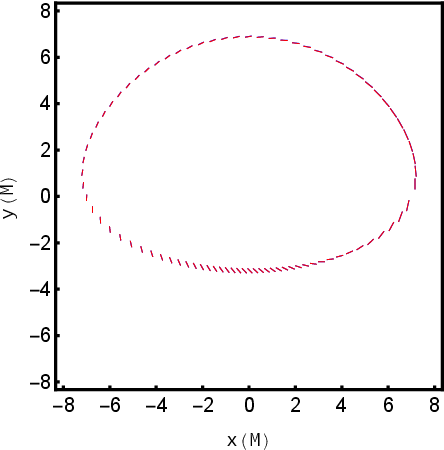}
\includegraphics[width=.24\linewidth]{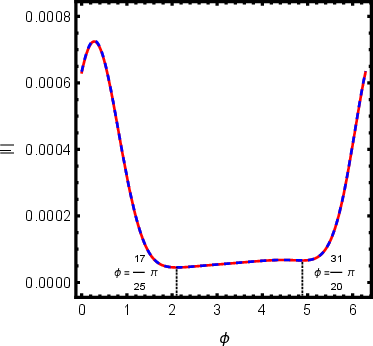} 
\includegraphics[width=.24\linewidth]{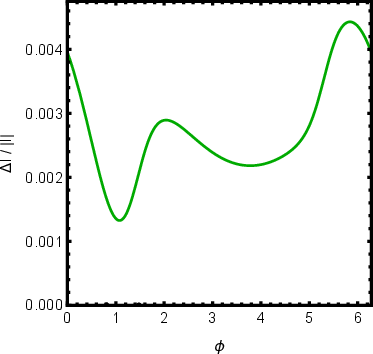} 
\includegraphics[width=.24\linewidth]{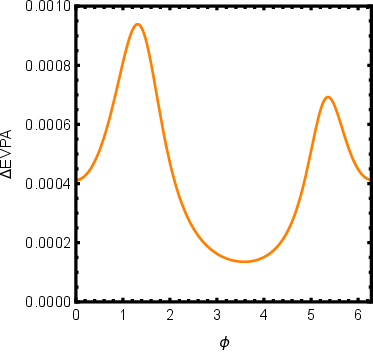}
\includegraphics[width=.24\linewidth]{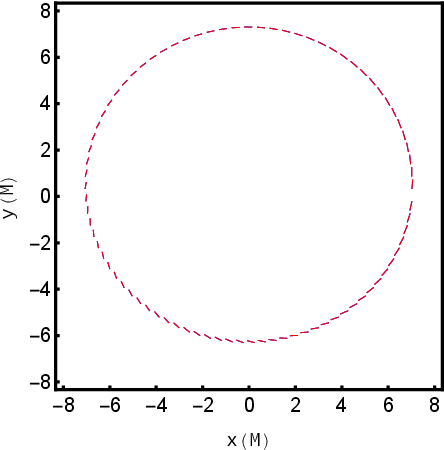}
\includegraphics[width=.24\linewidth]{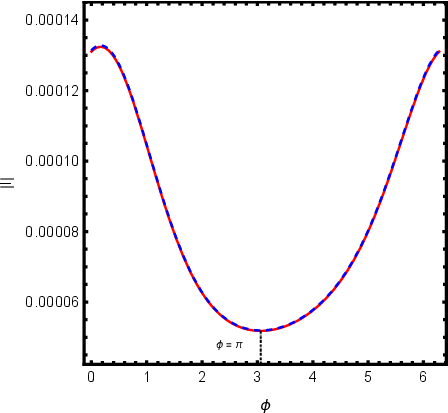}  
\includegraphics[width=.24\linewidth]{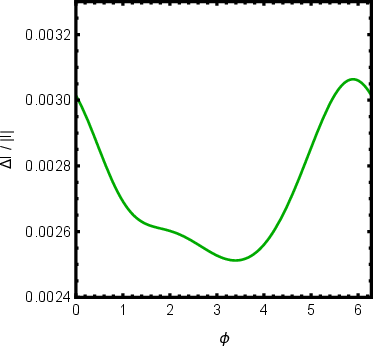} 
\includegraphics[width=.24\linewidth]{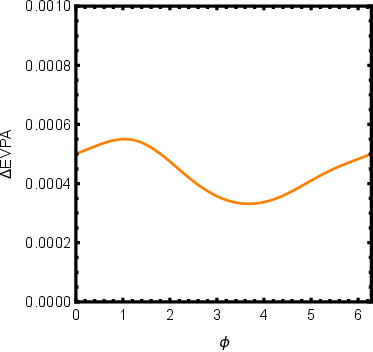}
\caption{ The same as Fig. \ref{equator_pic}, but for the values $q=0.5M$, $r_{s}=6M$. }
\label{equator_pic_q_05}
\end{figure*}

\section{setup}\label{sec-setup}

Now we are ready to plot polarization image and intensity for synchrotron radiation around regular black hole in general relativity coupled to nonlinear electrodynamics. We compare our results with the ones in the magnetically charged Reissner–Nordstr\"{o}m black hole in general relativity coupled to linear  electrodynamics. In our comparative analysis we assume the emitter moves along purely circular orbit confined to $\chi=-\pi/2$ around black hole in $\phi-$ direction with relativistic Keplerian orbital velocity given by
\begin{equation}
    \beta=\Omega \frac{r_{s}}{\sqrt{f(r_{s})}}\ ,
    \label{orb_vel}
\end{equation}
where orbital velocity $\beta$ of thin disk is measured in the local nonrotating frame. In the case of weakly charged thin disk, where specific charge of the disk is negligible, angular velocity is expressed as
\begin{equation}
    \Omega=\sqrt{-\frac{g_{tt,r}}{g_{\phi\phi,r}}}=\sqrt{\frac{f'(r_{s})}{2r_{s}}}\ ,
\end{equation}
where prime denotes derivative with respect to radial coordinate.
\begin{figure*}[t!]
\centering
\includegraphics[width=.33\linewidth]{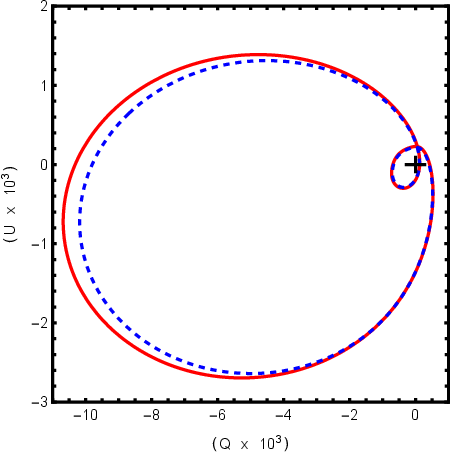}
\includegraphics[width=.33\linewidth]{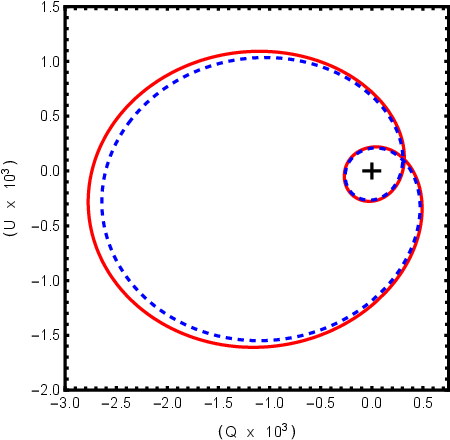} 
\includegraphics[width=.33\linewidth]{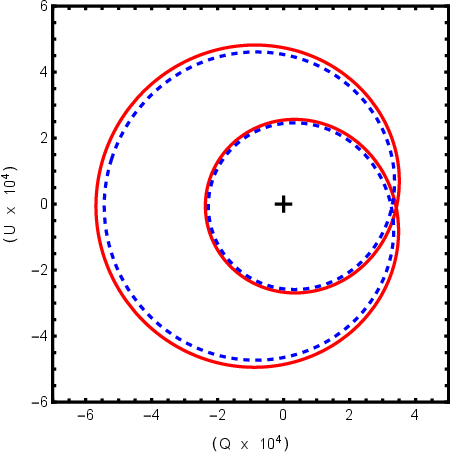}
\includegraphics[width=.33\linewidth]{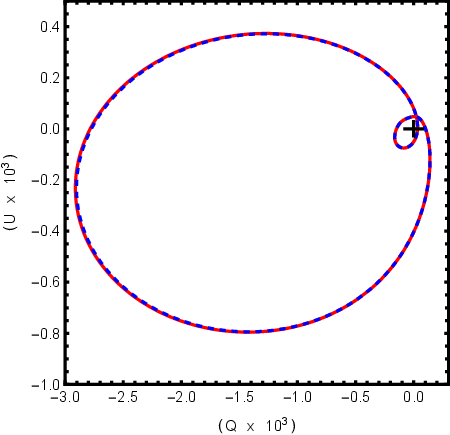}
\includegraphics[width=.33\linewidth]{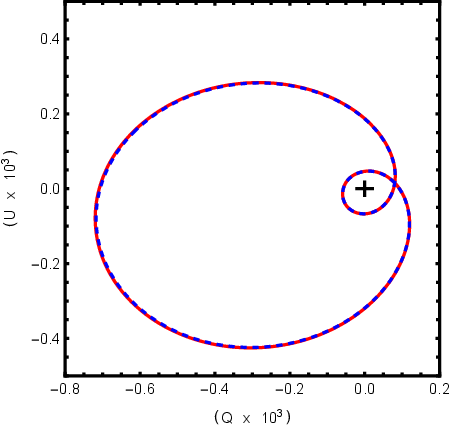} 
\includegraphics[width=.33\linewidth]{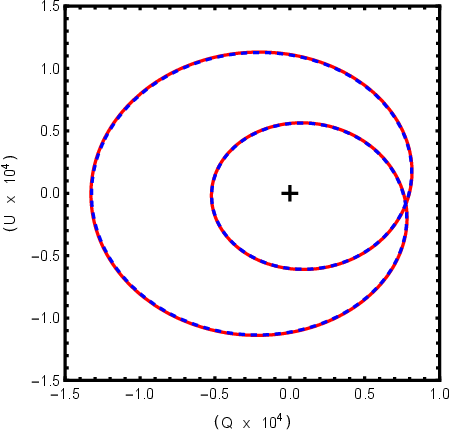}
\caption{Stokes parameters (Q-U diagrams) for $q=0.99M$ (top panel) and $q=0.50M$ (bottom panel) for different inclination angles $\theta_{0}=80^{\circ}, 60^{\circ}, 20^{\circ}$ (from left to right). The crosshair marks the origin of the axes.}
\label{Stokes_pic_q_0_99}
\end{figure*}

\section{Discussion}\label{sec-conc}

In the current paper we have studied the polarized image of a synchrotron emitting ring around spherically symmetric black holes. Specifically, we have chosen the regular Bronnikov black hole of general relativity coupled to nonlinear electrodynamics and the singular Reissner-Nordstr\"{o}m black hole of general relativity coupled to Maxwell electrodynamics, considering both black holes are magnetically charged. We have investigated the characteristics of synchrotron radiation in the vicinity of both black holes, by taking into account the fact that the photon does not follow the null geodesics of the original (Bronnikov black hole in the current case) spacetime in nonlinear electrodynamics; instead, it follows the null geodesics of the effective spacetime. The results were compared with the established solutions of the Reissner-Nordstr\"{o}m black hole with a magnetic charge, as shown in Figs. \ref{equator_pic}, \ref{equator_pic_q_099_rs_4_5}, \ref{equator_pic_q_05} and \ref{Stokes_pic_q_0_99}. In our comparisons, we used two values of magnetic charge, namely $q=0.5M$, $q=0.99M$ and two values of the ring radius $r_s=6M$, $r_s=4.5M$ in different inclination angles (ring tilt). It was assumed that the emitter moves in a circular orbit with $\cos\chi = -1$ and a given orbital velocity (Doppler and aberration) defined by the relation \eqref{orb_vel}. The overall results demonstrate that the intensity difference between the Bronnikov and Reissner-Nordstr\"{o}m black holes is significant at high magnetic charge and large tilt angles. Fig. \ref{equator_pic_q_099_rs_4_5} shows the polarization image for near extreme values of $r_{s}=4.5M$ which is close to the ISCO and magnetic charge $q=0.99M$, representing the closest equatorial ring for an almost extreme black hole. A comparison of the radiation intensities at $r_{s}=4.5M$ and $r_{s}=6M$ for $q=0.99M$ shows that the radiation intensity at $r_{s}=4.5M$ is about three times greater. For large tilt angles, the maximum intensity is located at approximately $\phi\approx 2\pi/25 $ and $\phi\approx \pi/10 $ for $q=0.99M$ and $q=0.5M$, respectively. Moreover, there exist two specific points where the radiation intensity becomes indistinguishable, as indicated by $\Delta I=0$, a phenomenon absent at smaller inclination angles (refer to the intensity difference observed for $\theta=80^{\circ}$). It is observed that, for identical tilt axes, the peak intensities of both the regular Bronnikov and the singular Reissner-Nordstr\"{o}m black holes are approximately 3.5 to 4.5 times greater at $q = 0.99M$ in comparison to $q = 0.5M$.

From the perspective of the electric vector polarization angle $\Delta$EVPA and normalized intensity difference $\Delta I/I$, it is clear that the morphological behaviors are nearly identical for the two different charge parameters. The significant differences in lensing and intensity at high magnetic charges arise from the distinct behavior of light in the effective metric compared to the standard metric. Consequently, the polarization observed in nonlinear electrodynamics coupled to general relativity is primarily influenced by factors such as the inverse emission radius (lensing), the Doppler effect, and redshift.

\section*{Acknowledgement}
A.T. and J.K. acknowledge the institutional support of the Research Centre for Theoretical Physics and Astrophysics, Institute of Physics, Silesian University in Opava. The work was also supported by the internal grant SGS/24/2024 of the Silesian University in Opava. A.T. also acknowledges the science and research grant in the Moravian-Silesian Region, Grant No.~RRC/09/2023 and thanks the ERASMUS+ ICM project for supporting his stay at the National University of Uzbekistan and extends his gratitude to Professor Bobomurat Ahmedov for his hospitality.

\bibliography{references}

\begin{thebibliography}{36}%
\makeatletter
\providecommand \@ifxundefined [1]{%
 \@ifx{#1\undefined}
}%
\providecommand \@ifnum [1]{%
 \ifnum #1\expandafter \@firstoftwo
 \else \expandafter \@secondoftwo
 \fi
}%
\providecommand \@ifx [1]{%
 \ifx #1\expandafter \@firstoftwo
 \else \expandafter \@secondoftwo
 \fi
}%
\providecommand \natexlab [1]{#1}%
\providecommand \enquote  [1]{``#1''}%
\providecommand \bibnamefont  [1]{#1}%
\providecommand \bibfnamefont [1]{#1}%
\providecommand \citenamefont [1]{#1}%
\providecommand \href@noop [0]{\@secondoftwo}%
\providecommand \href [0]{\begingroup \@sanitize@url \@href}%
\providecommand \@href[1]{\@@startlink{#1}\@@href}%
\providecommand \@@href[1]{\endgroup#1\@@endlink}%
\providecommand \@sanitize@url [0]{\catcode `\\12\catcode `\$12\catcode `\&12\catcode `\#12\catcode `\^12\catcode `\_12\catcode `\%12\relax}%
\providecommand \@@startlink[1]{}%
\providecommand \@@endlink[0]{}%
\providecommand \url  [0]{\begingroup\@sanitize@url \@url }%
\providecommand \@url [1]{\endgroup\@href {#1}{\urlprefix }}%
\providecommand \urlprefix  [0]{URL }%
\providecommand \Eprint [0]{\href }%
\providecommand \doibase [0]{http://dx.doi.org/}%
\providecommand \selectlanguage [0]{\@gobble}%
\providecommand \bibinfo  [0]{\@secondoftwo}%
\providecommand \bibfield  [0]{\@secondoftwo}%
\providecommand \translation [1]{[#1]}%
\providecommand \BibitemOpen [0]{}%
\providecommand \bibitemStop [0]{}%
\providecommand \bibitemNoStop [0]{.\EOS\space}%
\providecommand \EOS [0]{\spacefactor3000\relax}%
\providecommand \BibitemShut  [1]{\csname bibitem#1\endcsname}%
\let\auto@bib@innerbib\@empty
\bibitem [{\citenamefont {Akiyama}\ \emph {et~al.}(2019)\citenamefont {Akiyama} \emph {et~al.}}]{EHT:2019dse}%
  \BibitemOpen
  \bibfield  {author} {\bibinfo {author} {\bibfnamefont {K.}~\bibnamefont {Akiyama}} \emph {et~al.} (\bibinfo {collaboration} {Event Horizon Telescope}),\ }\href {\doibase 10.3847/2041-8213/ab0ec7} {\bibfield  {journal} {\bibinfo  {journal} {Astrophys. J. Lett.}\ }\textbf {\bibinfo {volume} {875}},\ \bibinfo {pages} {L1} (\bibinfo {year} {2019})},\ \Eprint {http://arxiv.org/abs/1906.11238} {arXiv:1906.11238 [astro-ph.GA]} \BibitemShut {NoStop}%
\bibitem [{\citenamefont {Akiyama}\ \emph {et~al.}(2022)\citenamefont {Akiyama} \emph {et~al.}}]{EHT:2022wkp}%
  \BibitemOpen
  \bibfield  {author} {\bibinfo {author} {\bibfnamefont {K.}~\bibnamefont {Akiyama}} \emph {et~al.} (\bibinfo {collaboration} {Event Horizon Telescope}),\ }\href {\doibase 10.3847/2041-8213/ac6674} {\bibfield  {journal} {\bibinfo  {journal} {Astrophys. J. Lett.}\ }\textbf {\bibinfo {volume} {930}},\ \bibinfo {pages} {L12} (\bibinfo {year} {2022})},\ \Eprint {http://arxiv.org/abs/2311.08680} {arXiv:2311.08680 [astro-ph.HE]} \BibitemShut {NoStop}%
\bibitem [{\citenamefont {Baub\"ock}\ \emph {et~al.}(2020)\citenamefont {Baub\"ock} \emph {et~al.}}]{GRAVITY:2020lpa}%
  \BibitemOpen
  \bibfield  {author} {\bibinfo {author} {\bibfnamefont {M.}~\bibnamefont {Baub\"ock}} \emph {et~al.} (\bibinfo {collaboration} {GRAVITY}),\ }\href {\doibase 10.1051/0004-6361/201937233} {\bibfield  {journal} {\bibinfo  {journal} {Astron. Astrophys.}\ }\textbf {\bibinfo {volume} {635}},\ \bibinfo {pages} {A143} (\bibinfo {year} {2020})},\ \Eprint {http://arxiv.org/abs/2002.08374} {arXiv:2002.08374 [astro-ph.HE]} \BibitemShut {NoStop}%
\bibitem [{\citenamefont {Akiyama}\ \emph {et~al.}(2021)\citenamefont {Akiyama} \emph {et~al.}}]{EHT:2021btj}%
  \BibitemOpen
  \bibfield  {author} {\bibinfo {author} {\bibfnamefont {K.}~\bibnamefont {Akiyama}} \emph {et~al.} (\bibinfo {collaboration} {Event Horizon Telescope}),\ }\href {\doibase 10.3847/1538-4357/abf117} {\bibfield  {journal} {\bibinfo  {journal} {Astrophys. J.}\ }\textbf {\bibinfo {volume} {912}},\ \bibinfo {pages} {35} (\bibinfo {year} {2021})},\ \Eprint {http://arxiv.org/abs/2105.01804} {arXiv:2105.01804 [astro-ph.HE]} \BibitemShut {NoStop}%
\bibitem [{\citenamefont {{Gelles}}\ \emph {et~al.}(2021)\citenamefont {{Gelles}}, \citenamefont {{Himwich}}, \citenamefont {{Johnson}},\ and\ \citenamefont {{Palumbo}}}]{2021PhRvD.104d4060G}%
  \BibitemOpen
  \bibfield  {author} {\bibinfo {author} {\bibfnamefont {Z.}~\bibnamefont {{Gelles}}}, \bibinfo {author} {\bibfnamefont {E.}~\bibnamefont {{Himwich}}}, \bibinfo {author} {\bibfnamefont {M.~D.}\ \bibnamefont {{Johnson}}}, \ and\ \bibinfo {author} {\bibfnamefont {D.~C.~M.}\ \bibnamefont {{Palumbo}}},\ }\href {\doibase 10.1103/PhysRevD.104.044060} {\bibfield  {journal} {\bibinfo  {journal} {\prd}\ }\textbf {\bibinfo {volume} {104}},\ \bibinfo {eid} {044060} (\bibinfo {year} {2021})},\ \Eprint {http://arxiv.org/abs/2105.09440} {arXiv:2105.09440 [gr-qc]} \BibitemShut {NoStop}%
\bibitem [{\citenamefont {{Zhang}}\ \emph {et~al.}(2022)\citenamefont {{Zhang}}, \citenamefont {{Chen}},\ and\ \citenamefont {{Jing}}}]{2022EPJC...82..835Z}%
  \BibitemOpen
  \bibfield  {author} {\bibinfo {author} {\bibfnamefont {Z.}~\bibnamefont {{Zhang}}}, \bibinfo {author} {\bibfnamefont {S.}~\bibnamefont {{Chen}}}, \ and\ \bibinfo {author} {\bibfnamefont {J.}~\bibnamefont {{Jing}}},\ }\href {\doibase 10.1140/epjc/s10052-022-10794-z} {\bibfield  {journal} {\bibinfo  {journal} {European Physical Journal C}\ }\textbf {\bibinfo {volume} {82}},\ \bibinfo {eid} {835} (\bibinfo {year} {2022})},\ \Eprint {http://arxiv.org/abs/2205.13696} {arXiv:2205.13696 [gr-qc]} \BibitemShut {NoStop}%
\bibitem [{\citenamefont {{Genzel}}\ \emph {et~al.}(2003)\citenamefont {{Genzel}}, \citenamefont {{Sch{\"o}del}}, \citenamefont {{Ott}}, \citenamefont {{Eckart}}, \citenamefont {{Alexander}}, \citenamefont {{Lacombe}}, \citenamefont {{Rouan}},\ and\ \citenamefont {{Aschenbach}}}]{2003Natur.425..934G}%
  \BibitemOpen
  \bibfield  {author} {\bibinfo {author} {\bibfnamefont {R.}~\bibnamefont {{Genzel}}}, \bibinfo {author} {\bibfnamefont {R.}~\bibnamefont {{Sch{\"o}del}}}, \bibinfo {author} {\bibfnamefont {T.}~\bibnamefont {{Ott}}}, \bibinfo {author} {\bibfnamefont {A.}~\bibnamefont {{Eckart}}}, \bibinfo {author} {\bibfnamefont {T.}~\bibnamefont {{Alexander}}}, \bibinfo {author} {\bibfnamefont {F.}~\bibnamefont {{Lacombe}}}, \bibinfo {author} {\bibfnamefont {D.}~\bibnamefont {{Rouan}}}, \ and\ \bibinfo {author} {\bibfnamefont {B.}~\bibnamefont {{Aschenbach}}},\ }\href {\doibase 10.1038/nature02065} {\bibfield  {journal} {\bibinfo  {journal} {\nat}\ }\textbf {\bibinfo {volume} {425}},\ \bibinfo {pages} {934} (\bibinfo {year} {2003})},\ \Eprint {http://arxiv.org/abs/astro-ph/0310821} {arXiv:astro-ph/0310821 [astro-ph]} \BibitemShut {NoStop}%
\bibitem [{\citenamefont {Will}(2014)}]{Will:2014kxa}%
  \BibitemOpen
  \bibfield  {author} {\bibinfo {author} {\bibfnamefont {C.~M.}\ \bibnamefont {Will}},\ }\href {\doibase 10.12942/lrr-2014-4} {\bibfield  {journal} {\bibinfo  {journal} {Living Rev. Rel.}\ }\textbf {\bibinfo {volume} {17}},\ \bibinfo {pages} {4} (\bibinfo {year} {2014})},\ \Eprint {http://arxiv.org/abs/1403.7377} {arXiv:1403.7377 [gr-qc]} \BibitemShut {NoStop}%
\bibitem [{\citenamefont {{The LIGO Scientific Collaboration and the Virgo Collaboration}}(2016)}]{GW151226}%
  \BibitemOpen
  \bibfield  {author} {\bibinfo {author} {\bibnamefont {{The LIGO Scientific Collaboration and the Virgo Collaboration}}},\ }\href {\doibase 10.1103/PhysRevLett.116.241103} {\bibfield  {journal} {\bibinfo  {journal} {Phys. Rev. Lett.}\ }\textbf {\bibinfo {volume} {116}},\ \bibinfo {eid} {241103} (\bibinfo {year} {2016})},\ \Eprint {http://arxiv.org/abs/1606.04855} {arXiv:1606.04855 [gr-qc]} \BibitemShut {NoStop}%
\bibitem [{\citenamefont {{The LIGO Scientific Collaboration and the Virgo Collaboration}}(2017{\natexlab{a}})}]{GW170104}%
  \BibitemOpen
  \bibfield  {author} {\bibinfo {author} {\bibnamefont {{The LIGO Scientific Collaboration and the Virgo Collaboration}}},\ }\href {\doibase 10.1103/PhysRevLett.118.221101} {\bibfield  {journal} {\bibinfo  {journal} {Phys. Rev. Lett.}\ }\textbf {\bibinfo {volume} {118}},\ \bibinfo {eid} {221101} (\bibinfo {year} {2017}{\natexlab{a}})},\ \Eprint {http://arxiv.org/abs/1706.01812} {arXiv:1706.01812 [gr-qc]} \BibitemShut {NoStop}%
\bibitem [{\citenamefont {{The LIGO Scientific Collaboration and the Virgo Collaboration}}(2017{\natexlab{b}})}]{GW170608}%
  \BibitemOpen
  \bibfield  {author} {\bibinfo {author} {\bibnamefont {{The LIGO Scientific Collaboration and the Virgo Collaboration}}},\ }\href {\doibase 10.3847/2041-8213/aa9f0c} {\bibfield  {journal} {\bibinfo  {journal} {Astrophys. J. Lett.}\ }\textbf {\bibinfo {volume} {851}},\ \bibinfo {eid} {L35} (\bibinfo {year} {2017}{\natexlab{b}})},\ \Eprint {http://arxiv.org/abs/1711.05578} {arXiv:1711.05578 [astro-ph.HE]} \BibitemShut {NoStop}%
\bibitem [{\citenamefont {{The LIGO Scientific Collaboration and the Virgo Collaboration}}(2017{\natexlab{c}})}]{GW170814}%
  \BibitemOpen
  \bibfield  {author} {\bibinfo {author} {\bibnamefont {{The LIGO Scientific Collaboration and the Virgo Collaboration}}},\ }\href {\doibase 10.1103/PhysRevLett.119.141101} {\bibfield  {journal} {\bibinfo  {journal} {Phys. Rev. Lett.}\ }\textbf {\bibinfo {volume} {119}},\ \bibinfo {eid} {141101} (\bibinfo {year} {2017}{\natexlab{c}})},\ \Eprint {http://arxiv.org/abs/1709.09660} {arXiv:1709.09660 [gr-qc]} \BibitemShut {NoStop}%
\bibitem [{\citenamefont {{The LIGO Scientific Collaboration and the Virgo Collaboration}}(2017{\natexlab{d}})}]{GW170817}%
  \BibitemOpen
  \bibfield  {author} {\bibinfo {author} {\bibnamefont {{The LIGO Scientific Collaboration and the Virgo Collaboration}}},\ }\href {\doibase 10.1103/PhysRevLett.119.161101} {\bibfield  {journal} {\bibinfo  {journal} {Phys. Rev. Lett.}\ }\textbf {\bibinfo {volume} {119}},\ \bibinfo {eid} {161101} (\bibinfo {year} {2017}{\natexlab{d}})},\ \Eprint {http://arxiv.org/abs/1710.05832} {arXiv:1710.05832 [gr-qc]} \BibitemShut {NoStop}%
\bibitem [{\citenamefont {Ayon-Beato}\ and\ \citenamefont {Garcia}(1998)}]{Ayon-Beato:1998hmi}%
  \BibitemOpen
  \bibfield  {author} {\bibinfo {author} {\bibfnamefont {E.}~\bibnamefont {Ayon-Beato}}\ and\ \bibinfo {author} {\bibfnamefont {A.}~\bibnamefont {Garcia}},\ }\href {\doibase 10.1103/PhysRevLett.80.5056} {\bibfield  {journal} {\bibinfo  {journal} {Phys. Rev. Lett.}\ }\textbf {\bibinfo {volume} {80}},\ \bibinfo {pages} {5056} (\bibinfo {year} {1998})},\ \Eprint {http://arxiv.org/abs/gr-qc/9911046} {arXiv:gr-qc/9911046} \BibitemShut {NoStop}%
\bibitem [{\citenamefont {Bronnikov}(2000)}]{Bronnikov:2000yz}%
  \BibitemOpen
  \bibfield  {author} {\bibinfo {author} {\bibfnamefont {K.~A.}\ \bibnamefont {Bronnikov}},\ }\href {\doibase 10.1103/PhysRevLett.85.4641} {\bibfield  {journal} {\bibinfo  {journal} {Phys. Rev. Lett.}\ }\textbf {\bibinfo {volume} {85}},\ \bibinfo {pages} {4641} (\bibinfo {year} {2000})}\BibitemShut {NoStop}%
\bibitem [{\citenamefont {{Dymnikova}}(2004)}]{Dymnikova04}%
  \BibitemOpen
  \bibfield  {author} {\bibinfo {author} {\bibfnamefont {I.}~\bibnamefont {{Dymnikova}}},\ }\href {\doibase 10.1088/0264-9381/21/18/009} {\bibfield  {journal} {\bibinfo  {journal} {Classical and Quantum Gravity}\ }\textbf {\bibinfo {volume} {21}},\ \bibinfo {pages} {4417} (\bibinfo {year} {2004})},\ \Eprint {http://arxiv.org/abs/gr-qc/0407072} {gr-qc/0407072} \BibitemShut {NoStop}%
\bibitem [{\citenamefont {{Novello}}\ \emph {et~al.}(2001)\citenamefont {{Novello}}, \citenamefont {{Salim}}, \citenamefont {{De Lorenci}},\ and\ \citenamefont {{Elbaz}}}]{Novello01}%
  \BibitemOpen
  \bibfield  {author} {\bibinfo {author} {\bibfnamefont {M.}~\bibnamefont {{Novello}}}, \bibinfo {author} {\bibfnamefont {J.~M.}\ \bibnamefont {{Salim}}}, \bibinfo {author} {\bibfnamefont {V.~A.}\ \bibnamefont {{De Lorenci}}}, \ and\ \bibinfo {author} {\bibfnamefont {E.}~\bibnamefont {{Elbaz}}},\ }\href {\doibase 10.1103/PhysRevD.63.103516} {\bibfield  {journal} {\bibinfo  {journal} {Phys. Rev. D}\ }\textbf {\bibinfo {volume} {63}},\ \bibinfo {eid} {103516} (\bibinfo {year} {2001})}\BibitemShut {NoStop}%
\bibitem [{\citenamefont {{Obukhov}}\ and\ \citenamefont {{Rubilar}}(2002)}]{Obukhov02}%
  \BibitemOpen
  \bibfield  {author} {\bibinfo {author} {\bibfnamefont {Y.~N.}\ \bibnamefont {{Obukhov}}}\ and\ \bibinfo {author} {\bibfnamefont {G.~F.}\ \bibnamefont {{Rubilar}}},\ }\href {\doibase 10.1103/PhysRevD.66.024042} {\bibfield  {journal} {\bibinfo  {journal} {Phys. Rev. D}\ }\textbf {\bibinfo {volume} {66}},\ \bibinfo {eid} {024042} (\bibinfo {year} {2002})},\ \Eprint {http://arxiv.org/abs/gr-qc/0204028} {gr-qc/0204028} \BibitemShut {NoStop}%
\bibitem [{\citenamefont {{Lemos}}\ and\ \citenamefont {{Zanchin}}(2016)}]{Lemos16}%
  \BibitemOpen
  \bibfield  {author} {\bibinfo {author} {\bibfnamefont {J.~P.~S.}\ \bibnamefont {{Lemos}}}\ and\ \bibinfo {author} {\bibfnamefont {V.~T.}\ \bibnamefont {{Zanchin}}},\ }\href {\doibase 10.1103/PhysRevD.93.124012} {\bibfield  {journal} {\bibinfo  {journal} {Phys. Rev. D}\ }\textbf {\bibinfo {volume} {93}},\ \bibinfo {eid} {124012} (\bibinfo {year} {2016})},\ \Eprint {http://arxiv.org/abs/1603.07359} {arXiv:1603.07359 [gr-qc]} \BibitemShut {NoStop}%
\bibitem [{\citenamefont {{Bret{\'o}n}}\ and\ \citenamefont {{L{\'o}pez}}(2016)}]{Breto16}%
  \BibitemOpen
  \bibfield  {author} {\bibinfo {author} {\bibfnamefont {N.}~\bibnamefont {{Bret{\'o}n}}}\ and\ \bibinfo {author} {\bibfnamefont {L.~A.}\ \bibnamefont {{L{\'o}pez}}},\ }\href {\doibase 10.1103/PhysRevD.94.104008} {\bibfield  {journal} {\bibinfo  {journal} {Phys. Rev. D}\ }\textbf {\bibinfo {volume} {94}},\ \bibinfo {eid} {104008} (\bibinfo {year} {2016})},\ \Eprint {http://arxiv.org/abs/1607.02476} {arXiv:1607.02476 [gr-qc]} \BibitemShut {NoStop}%
\bibitem [{\citenamefont {{Fan}}\ and\ \citenamefont {{Wang}}(2016)}]{Fan:PRD:2016}%
  \BibitemOpen
  \bibfield  {author} {\bibinfo {author} {\bibfnamefont {Z.-Y.}\ \bibnamefont {{Fan}}}\ and\ \bibinfo {author} {\bibfnamefont {X.}~\bibnamefont {{Wang}}},\ }\href {\doibase 10.1103/PhysRevD.94.124027} {\bibfield  {journal} {\bibinfo  {journal} {Phys. Rev. D}\ }\textbf {\bibinfo {volume} {94}},\ \bibinfo {eid} {124027} (\bibinfo {year} {2016})},\ \Eprint {http://arxiv.org/abs/1610.02636} {arXiv:1610.02636 [gr-qc]} \BibitemShut {NoStop}%
\bibitem [{\citenamefont {Toshmatov}\ \emph {et~al.}(2018)\citenamefont {Toshmatov}, \citenamefont {Stuchl\'\i{}k},\ and\ \citenamefont {Ahmedov}}]{Toshmatov:2018ell}%
  \BibitemOpen
  \bibfield  {author} {\bibinfo {author} {\bibfnamefont {B.}~\bibnamefont {Toshmatov}}, \bibinfo {author} {\bibfnamefont {Z.}~\bibnamefont {Stuchl\'\i{}k}}, \ and\ \bibinfo {author} {\bibfnamefont {B.}~\bibnamefont {Ahmedov}},\ }\href {\doibase 10.1103/PhysRevD.98.085021} {\bibfield  {journal} {\bibinfo  {journal} {Phys. Rev. D}\ }\textbf {\bibinfo {volume} {98}},\ \bibinfo {pages} {085021} (\bibinfo {year} {2018})},\ \Eprint {http://arxiv.org/abs/1810.06383} {arXiv:1810.06383 [gr-qc]} \BibitemShut {NoStop}%
\bibitem [{\citenamefont {Nomura}\ \emph {et~al.}(2020)\citenamefont {Nomura}, \citenamefont {Yoshida},\ and\ \citenamefont {Soda}}]{Nomura:2020tpc}%
  \BibitemOpen
  \bibfield  {author} {\bibinfo {author} {\bibfnamefont {K.}~\bibnamefont {Nomura}}, \bibinfo {author} {\bibfnamefont {D.}~\bibnamefont {Yoshida}}, \ and\ \bibinfo {author} {\bibfnamefont {J.}~\bibnamefont {Soda}},\ }\href {\doibase 10.1103/PhysRevD.101.124026} {\bibfield  {journal} {\bibinfo  {journal} {Phys. Rev. D}\ }\textbf {\bibinfo {volume} {101}},\ \bibinfo {pages} {124026} (\bibinfo {year} {2020})},\ \Eprint {http://arxiv.org/abs/2004.07560} {arXiv:2004.07560 [gr-qc]} \BibitemShut {NoStop}%
\bibitem [{\citenamefont {Toshmatov}\ \emph {et~al.}(2021)\citenamefont {Toshmatov}, \citenamefont {Ahmedov},\ and\ \citenamefont {Malafarina}}]{Toshmatov:2021fgm}%
  \BibitemOpen
  \bibfield  {author} {\bibinfo {author} {\bibfnamefont {B.}~\bibnamefont {Toshmatov}}, \bibinfo {author} {\bibfnamefont {B.}~\bibnamefont {Ahmedov}}, \ and\ \bibinfo {author} {\bibfnamefont {D.}~\bibnamefont {Malafarina}},\ }\href {\doibase 10.1103/PhysRevD.103.024026} {\bibfield  {journal} {\bibinfo  {journal} {Phys. Rev. D}\ }\textbf {\bibinfo {volume} {103}},\ \bibinfo {pages} {024026} (\bibinfo {year} {2021})},\ \Eprint {http://arxiv.org/abs/2101.05496} {arXiv:2101.05496 [gr-qc]} \BibitemShut {NoStop}%
\bibitem [{\citenamefont {Moreno}\ and\ \citenamefont {Sarbach}(2003)}]{Moreno:2002gg}%
  \BibitemOpen
  \bibfield  {author} {\bibinfo {author} {\bibfnamefont {C.}~\bibnamefont {Moreno}}\ and\ \bibinfo {author} {\bibfnamefont {O.}~\bibnamefont {Sarbach}},\ }\href {\doibase 10.1103/PhysRevD.67.024028} {\bibfield  {journal} {\bibinfo  {journal} {Phys. Rev. D}\ }\textbf {\bibinfo {volume} {67}},\ \bibinfo {pages} {024028} (\bibinfo {year} {2003})},\ \Eprint {http://arxiv.org/abs/gr-qc/0208090} {arXiv:gr-qc/0208090} \BibitemShut {NoStop}%
\bibitem [{\citenamefont {{Schee}}\ and\ \citenamefont {{Stuchl{\'\i}k}}(2019)}]{2019ApJ87412S}%
  \BibitemOpen
  \bibfield  {author} {\bibinfo {author} {\bibfnamefont {J.}~\bibnamefont {{Schee}}}\ and\ \bibinfo {author} {\bibfnamefont {Z.}~\bibnamefont {{Stuchl{\'\i}k}}},\ }\href {\doibase 10.3847/1538-4357/ab04f3} {\bibfield  {journal} {\bibinfo  {journal} {The Astrophysical Journal}\ }\textbf {\bibinfo {volume} {874}},\ \bibinfo {eid} {12} (\bibinfo {year} {2019})}\BibitemShut {NoStop}%
\bibitem [{\citenamefont {Toshmatov}\ \emph {et~al.}(2019)\citenamefont {Toshmatov}, \citenamefont {Stuchl\'\i{}k}, \citenamefont {Ahmedov},\ and\ \citenamefont {Malafarina}}]{Toshmatov:2019gxg}%
  \BibitemOpen
  \bibfield  {author} {\bibinfo {author} {\bibfnamefont {B.}~\bibnamefont {Toshmatov}}, \bibinfo {author} {\bibfnamefont {Z.}~\bibnamefont {Stuchl\'\i{}k}}, \bibinfo {author} {\bibfnamefont {B.}~\bibnamefont {Ahmedov}}, \ and\ \bibinfo {author} {\bibfnamefont {D.}~\bibnamefont {Malafarina}},\ }\href {\doibase 10.1103/PhysRevD.99.064043} {\bibfield  {journal} {\bibinfo  {journal} {Phys. Rev. D}\ }\textbf {\bibinfo {volume} {99}},\ \bibinfo {pages} {064043} (\bibinfo {year} {2019})},\ \Eprint {http://arxiv.org/abs/1903.03778} {arXiv:1903.03778 [gr-qc]} \BibitemShut {NoStop}%
\bibitem [{\citenamefont {{Stuchl{\'\i}k}}\ \emph {et~al.}(2019)\citenamefont {{Stuchl{\'\i}k}}, \citenamefont {{Schee}},\ and\ \citenamefont {{Ovchinnikov}}}]{2019ApJ145S}%
  \BibitemOpen
  \bibfield  {author} {\bibinfo {author} {\bibfnamefont {Z.}~\bibnamefont {{Stuchl{\'\i}k}}}, \bibinfo {author} {\bibfnamefont {J.}~\bibnamefont {{Schee}}}, \ and\ \bibinfo {author} {\bibfnamefont {D.}~\bibnamefont {{Ovchinnikov}}},\ }\href {\doibase 10.3847/1538-4357/ab55d5} {\bibfield  {journal} {\bibinfo  {journal} {The Astrophysical Journal}\ }\textbf {\bibinfo {volume} {887}},\ \bibinfo {eid} {145} (\bibinfo {year} {2019})}\BibitemShut {NoStop}%
\bibitem [{\citenamefont {{Liu}}\ \emph {et~al.}(2022)\citenamefont {{Liu}}, \citenamefont {{Chen}},\ and\ \citenamefont {{Jing}}}]{2022SCPMA..6520411L}%
  \BibitemOpen
  \bibfield  {author} {\bibinfo {author} {\bibfnamefont {X.}~\bibnamefont {{Liu}}}, \bibinfo {author} {\bibfnamefont {S.}~\bibnamefont {{Chen}}}, \ and\ \bibinfo {author} {\bibfnamefont {J.}~\bibnamefont {{Jing}}},\ }\href {\doibase 10.1007/s11433-022-1946-2} {\bibfield  {journal} {\bibinfo  {journal} {Science China Physics, Mechanics, and Astronomy}\ }\textbf {\bibinfo {volume} {65}},\ \bibinfo {eid} {120411} (\bibinfo {year} {2022})},\ \Eprint {http://arxiv.org/abs/2205.00391} {arXiv:2205.00391 [gr-qc]} \BibitemShut {NoStop}%
\bibitem [{\citenamefont {{Bronnikov}}(2001)}]{Bronnikov01}%
  \BibitemOpen
  \bibfield  {author} {\bibinfo {author} {\bibfnamefont {K.~A.}\ \bibnamefont {{Bronnikov}}},\ }\href {\doibase 10.1103/PhysRevD.63.044005} {\bibfield  {journal} {\bibinfo  {journal} {Phys. Rev. D}\ }\textbf {\bibinfo {volume} {63}},\ \bibinfo {eid} {044005} (\bibinfo {year} {2001})},\ \Eprint {http://arxiv.org/abs/gr-qc/0006014} {gr-qc/0006014} \BibitemShut {NoStop}%
\bibitem [{\citenamefont {Novello}\ \emph {et~al.}(2000{\natexlab{a}})\citenamefont {Novello}, \citenamefont {De~Lorenci}, \citenamefont {Salim},\ and\ \citenamefont {Klippert}}]{Novello:1999pg}%
  \BibitemOpen
  \bibfield  {author} {\bibinfo {author} {\bibfnamefont {M.}~\bibnamefont {Novello}}, \bibinfo {author} {\bibfnamefont {V.~A.}\ \bibnamefont {De~Lorenci}}, \bibinfo {author} {\bibfnamefont {J.~M.}\ \bibnamefont {Salim}}, \ and\ \bibinfo {author} {\bibfnamefont {R.}~\bibnamefont {Klippert}},\ }\href {\doibase 10.1103/PhysRevD.61.045001} {\bibfield  {journal} {\bibinfo  {journal} {Phys. Rev. D}\ }\textbf {\bibinfo {volume} {61}},\ \bibinfo {pages} {045001} (\bibinfo {year} {2000}{\natexlab{a}})},\ \Eprint {http://arxiv.org/abs/gr-qc/9911085} {arXiv:gr-qc/9911085} \BibitemShut {NoStop}%
\bibitem [{\citenamefont {Novello}\ \emph {et~al.}(2000{\natexlab{b}})\citenamefont {Novello}, \citenamefont {Perez~Bergliaffa},\ and\ \citenamefont {Salim}}]{Novello:2000km}%
  \BibitemOpen
  \bibfield  {author} {\bibinfo {author} {\bibfnamefont {M.}~\bibnamefont {Novello}}, \bibinfo {author} {\bibfnamefont {S.~E.}\ \bibnamefont {Perez~Bergliaffa}}, \ and\ \bibinfo {author} {\bibfnamefont {J.~M.}\ \bibnamefont {Salim}},\ }\href {\doibase 10.1088/0264-9381/17/18/316} {\bibfield  {journal} {\bibinfo  {journal} {Class. Quant. Grav.}\ }\textbf {\bibinfo {volume} {17}},\ \bibinfo {pages} {3821} (\bibinfo {year} {2000}{\natexlab{b}})},\ \Eprint {http://arxiv.org/abs/gr-qc/0003052} {arXiv:gr-qc/0003052} \BibitemShut {NoStop}%
\bibitem [{\citenamefont {{Novello}}\ \emph {et~al.}(2000)\citenamefont {{Novello}}, \citenamefont {{Perez Bergliaffa}},\ and\ \citenamefont {{Salim}}}]{2000CQGra..17.3821N}%
  \BibitemOpen
  \bibfield  {author} {\bibinfo {author} {\bibfnamefont {M.}~\bibnamefont {{Novello}}}, \bibinfo {author} {\bibfnamefont {S.~E.}\ \bibnamefont {{Perez Bergliaffa}}}, \ and\ \bibinfo {author} {\bibfnamefont {J.~M.}\ \bibnamefont {{Salim}}},\ }\href {\doibase 10.1088/0264-9381/17/18/316} {\bibfield  {journal} {\bibinfo  {journal} {Classical and Quantum Gravity}\ }\textbf {\bibinfo {volume} {17}},\ \bibinfo {pages} {3821} (\bibinfo {year} {2000})},\ \Eprint {http://arxiv.org/abs/gr-qc/0003052} {arXiv:gr-qc/0003052 [gr-qc]} \BibitemShut {NoStop}%
\bibitem [{\citenamefont {{Qin}}\ \emph {et~al.}(2022)\citenamefont {{Qin}}, \citenamefont {{Chen}},\ and\ \citenamefont {{Jing}}}]{2022EPJC...82..784Q}%
  \BibitemOpen
  \bibfield  {author} {\bibinfo {author} {\bibfnamefont {X.}~\bibnamefont {{Qin}}}, \bibinfo {author} {\bibfnamefont {S.}~\bibnamefont {{Chen}}}, \ and\ \bibinfo {author} {\bibfnamefont {J.}~\bibnamefont {{Jing}}},\ }\href {\doibase 10.1140/epjc/s10052-022-10753-8} {\bibfield  {journal} {\bibinfo  {journal} {European Physical Journal C}\ }\textbf {\bibinfo {volume} {82}},\ \bibinfo {eid} {784} (\bibinfo {year} {2022})},\ \Eprint {http://arxiv.org/abs/2111.10138} {arXiv:2111.10138 [gr-qc]} \BibitemShut {NoStop}%
\bibitem [{\citenamefont {{Beloborodov}}(2002)}]{2002ApJ...566L..85B}%
  \BibitemOpen
  \bibfield  {author} {\bibinfo {author} {\bibfnamefont {A.~M.}\ \bibnamefont {{Beloborodov}}},\ }\href {\doibase 10.1086/339511} {\bibfield  {journal} {\bibinfo  {journal} {\apjl}\ }\textbf {\bibinfo {volume} {566}},\ \bibinfo {pages} {L85} (\bibinfo {year} {2002})},\ \Eprint {http://arxiv.org/abs/astro-ph/0201117} {arXiv:astro-ph/0201117 [astro-ph]} \BibitemShut {NoStop}%
\bibitem [{\citenamefont {{Walker}}\ and\ \citenamefont {{Penrose}}(1970)}]{1970CMaPh..18..265W}%
  \BibitemOpen
  \bibfield  {author} {\bibinfo {author} {\bibfnamefont {M.}~\bibnamefont {{Walker}}}\ and\ \bibinfo {author} {\bibfnamefont {R.}~\bibnamefont {{Penrose}}},\ }\href {\doibase 10.1007/BF01649445} {\bibfield  {journal} {\bibinfo  {journal} {Communications in Mathematical Physics}\ }\textbf {\bibinfo {volume} {18}},\ \bibinfo {pages} {265} (\bibinfo {year} {1970})}\BibitemShut {NoStop}%
\end{thebibliography}%
\end{document}